    \documentclass[prl,aps,twocolumn,showpacs]{revtex4}
\usepackage[dvips]{graphicx}
\usepackage{dcolumn}%
\usepackage{amsmath}%
\usepackage{amsfonts}%
\usepackage{amssymb}
\providecommand{\U}[1]{\protect\rule{.1in}{.1in}}
\newcommand{\be}{\begin{equation}}
\newcommand{\ee}{\end{equation}}
\newcommand{\bea}{\begin{eqnarray}}
\newcommand{\eea}{\end{eqnarray}}
\newcommand{\nn}{ \nonumber}
\newcommand{\ds}{\displaystyle}
\begin{document}

\title{Vibration-induced inelastic effects in the electron transport through multisite molecular bridges}

\author{ Natalya A. Zimbovskaya$^{1}$ and Maija M. Kuklja$^2$}

\affiliation{$^1$Department of Physics and  Electronics, University of Puerto Rico, Humacao, PR 00791, \\ 
Institute for Functional Nanomaterials, University of Puerto Rico, San Juan, PR 00931, }
\affiliation{ 
 $^2$Materials Science and Engineering Department, University of Maryland, College Park, Maryland  20742-2115}

\begin{abstract} 
  We theoretically analyzed inelastic effects in the electron transport through molecular junctions originating from  electron-vibron interactions. The molecular bridge was simulated by a periodic chain of identical interacting hydrogen-like atoms  providing a set of energy states for the electron tunneling. To avoid difficulties inevitably arising when advanced computational techniques are employed to study inelastic electron transport through multilevel bridges, we propose and develop a semiphenomenological approach. The latter is based on the Buttiker's dephasing model within the scattering matrix formalism. The advantage of the proposed approach is that it allows to analytically study various inelastic effects on the conduction through molecular junctions including multilevel bridges. Here,  we apply this approach to describe features associated with electron energy transfer to vibrational phonons which appear in the inelastic tunneling spectra of electrons. In the particular case of a single level bridge our results agree with those obtained by self-consitent calculations carried out within the nonequilibrium Green's functions method validating the usefulness of the suggested approach.
 \end{abstract}


\date{\today}
\maketitle

\section{I. introduction}

Molecular electronics is currently recognized as one of the most promising developments in nanoelectronics, and a significant progress had been seen in this field during the past decade \cite{1,2}. The key element and basic building block of molecular electronic devices is a molecular junction consisting of two conducting leads linked by a molecular bridge. Conduction through molecular junctions is being intensively studied both theoretically and experimentally, and main transport mechanisms are presently understood \cite{2,3}. However, the progress of experimental capabilities in the field of molecular electronics unceasingly brings new challenges urging further development of the theory.

In practical molecular junctions electron transport is always accompanied by nuclear motions in the surroundings, so the conduction is affected by the coupling between electronic and vibrational degrees of freedom. To analyze vibration-induced effects in the electron transport through molecular bridges one assumes that molecular orbitals are coupled to phonons  describing the vibrations, and this brings   an inelastic contribution to the electron current.
  Over the past decade theoretical studies of inelastic vibration-induced effects in the electron transport through molecules and other similar nanosystems (e.g. carbon nanotubes) were carried out by a large number of authors (see Refs. \cite{4,5,6,7,8,9,10,11,12,13}). Also, manifestations of the electron-vibron interactions were observed in the experiments using various molecules, carbon nanotubes and fullerenes as linkers in the junctions \cite{14,15,16,17,18,19,20,21,22,23}. Vibration-induced features were viewed in the differential molecular conductance $ dI/dV \ (I,V$ respectively being the electron current through the junction, and the bias voltage applied across the latter) and in the inelastic tunneling spectra of electrons $d^2I/dV^2.$ Sometimes even the $I-V$ curves themselves exhibit an extra step originating from the electron-vibron interactions, as was reported in the Ref. \cite{17} where the electron transport through a single hydrogen molecule placed in between platinum leads was explored.

Here, we focus on theoretical studies of the inelastic electron tunneling spectra (IETS) which is the rich and diverse problem attracting an unceasing interest of the research community for a long while \cite{5,6,7,13,24,25}. A systematic framework for such studies is based on the nonequilibrium Green's functions formalism (NEGF). Within the NEGF the electron transport through a two terminal junction including a linking molecule (presented as a set of energy levels $E_i$) coupled to a set of phonons with the frequencies $\Omega_i$  could be properly analyzed. However, the extreme complexity  of the general equations for the relevant interdependent electron and phonon Green's functions limits the practical applicability of NEGF to simple models. Currently, specific NEGF based results for IETS are mostly obtained for a model where the molecule is simulated by a single orbital (a single-state bridge) which is coupled to a single vibrational mode with the frequency $ \Omega.$ Adopting this simple model, it was shown that a vibration-induced signal appears in the IETS near the phonon excitation threshold $ eV = \hbar \Omega.$ The signal looks as a peak, dip or derivative-like feature in the voltage dependence of $ d^2 I/dV^2. $ Its specific shape is  sensitive to the junction characteristics such as the electronic state energy, electron-phonon and molecule-to leads coupling strengths and the vibronic frequency (see the recent review and references therein \cite{7}).

In the present work we analyze the IETS using more realistic model to mimic a molecule linking the terminals in the junction. We simulate the molecule by a periodical chain of identical hydrogen-like atoms with the nearest neighbors interaction. The first and the last sites in chain are coupled to the leads. Such a model was fist proposed by D'Amato and Pastawski \cite{26}, and afterwards it was repeatedly used to study electron conduction through molecules. The chosen model gives us to describe a multisite bridge for the electron transport. The number of electronic energy states on the bridge equals to the number of the bridge sites \cite{27}, and the states are situated within the energy range whose width is determined by the coupling between the sites. To avoid computational difficulties arising when the consistent NEGF is employed, we chose a semiphenomenological approach based on the combination of the latter with the Buttiker dephasing model \cite{28}. This enables us to significantly reduce computational effort. Earlier, the similar approach was successfully applied to analytically analyze dissipative electron conduction through single site molecular bridges \cite{29}. Within the chosen approach we analyze the vibration-induced features  in the $d^2I/dV^2$ versus $ V$ spectrum, and we show that the spectrum may significantly differ from the spectra typical for single site molecular bridges. Besides the peak/dip or derivative-like feature at the phonon excitation threshold, extra IETS features may appear, whose location, size and shape strongly depend on the number of sites and site - to site couplings in the bridge.

\section{ii. model and main equations }

An important advantage of the phenomenological model for the incoherent/inelastic quantum transport proposed by Buttiker \cite{28} is that this model could be easily adapted to analyze various inelastic effects in the electron transport through molecular junctions. Within this model one may analytically study such effects avoiding inevitable technical difficulties which arise when the advanced methods are applied to compute characteristics of the electron transport through multisite (and multilevel) molecular bridges. Following Buttiker's approach, inelastic effects are accounted for by means of  electron reservoirs attached to the bridge sites. Electrons could be scattered into these reservoirs where they participate in dephasing/inelastic processes before their returning back to the bridge.  
 
 In further calculations, we mimic a molecular bridge linking two electrodes as a periodic chain of identical sites including $2m-1 $ elements. We set the site ionization energies $ E_i = E_0 $ and nearest neighbors coupling strengths $ \beta_{i,i+1} = \beta_{i,i-1} = \beta. $ 
 The schematic of the adopted model is shown in Fig. 1. 
  The first and the last sites in the chain are attached to the leads and remain fixed. We suggest that the chain vibrates at a fundamental frequency $\Omega ,$ and we study the electrons interaction with the corresponding transverse mode. In this case, the maximum displacement of the chain sites occurs in the middle of the latter. Therefore we assume, as a first approximation, that only the middle site in the chain is coupled to the vibrational mode. To describe this coupling  within the Buttiker approach, we put the middle site in contact with an electron reservoir. The electrons scattered in there could absorb/emit phonons and then reappear on the bridge.

\begin{figure}[t]  
\begin{center}
\includegraphics[width=4.5cm,height=9cm,angle=-90]{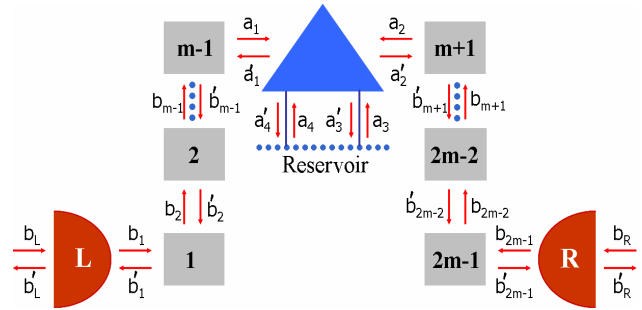}
\end{center}
\caption{(Color online) Schematic drawing illustrating  electron transport through the junction including a chain of $2m -1$ identical sites. The dissipative electron reservoir is attached to the middle site. The semicircles represent left $(L)$ and right $(R)$ electrodes. }%
\label{rateI}%
\end{figure}

 In the figure 1, the squares represent  barriers separating  sites included into the chain and the triangle imitates the scatterer coupling the middle site to the reservoir. The scatterer is described by the scattering matrix $s $ which relates outgoing from the scatterer wave amplitudes $a_1', a_2', a_3', a_4'$ to the incoming amplitudes  $a_1, a_2, a_3, a_4.$ The latter has the form \cite{28}:
   \be  
  s = 
\left( \begin{array}{cccc}
0 & \ \sqrt{1 - \epsilon} &\ \sqrt\epsilon &\ 0\\
 \sqrt{1 - \epsilon} & 0 & 0 & \sqrt\epsilon \\
\sqrt{\epsilon} & 0 & 0 & -\sqrt{1- \epsilon} \\
0 & \sqrt{\epsilon} &- \sqrt{1-\epsilon }& 0 \\
\end{array} \right) . \label{1}
  \ee
  Here, the parameter $\epsilon$ takes on values within the range $[0,1]. $ This parameter determines the probability for an electron to be scattered, and it was introduced in Ref. \cite{28} as a phenomenological characteristic.
  In further analysis, we suggest to express $\epsilon $ in terms of relevant energies characterizing electron transport through the molecular junction (see the next Section). This provides means to better describe  the transport processes within the  chosen scattering matrix formalism.

  To simplify the following calculations of the electron transmission, we assume that an electron could be injected at the system shown in Fig. 1 and/or leave from there solely via four channels indicated in the figure. Then incoming particle fluxes $ J_k $ are related to those outgoing from the junction $(J_j')$ by means of the $4\times 4$ transmission matrix $ T $ \cite{28,29}:
  \be
   J_j' = \sum_k T_{jk} J_k; \qquad j \leq  k \leq 4 ,    \label{2}
  \ee
  where $J_1,J_1' \equiv J_L, J_L' $ correspond to the left side of the junction and the fluxes  $J_2,J_2' \equiv J_R, J_R' $ are associated with its right side. To provide charge conservation in the system, the net particle flux in the channels $3,4 $ must equal zero.
 
   Assuming the electron transport from the left to the right,  
the electron transmission is the ratio of particle flux outgoing from the right end of the junction $2$ and that incoming to its left end $ 1.$ 
  To arrive at the expression for  $T(E), $ one must solve the system of linear equations (\ref{2}).   This gives:
      \be 
 T(E) = T_{21} + \frac{K_1(E) K_2(E)}{2 - R(E)}. \label{3}
  \ee
Here,
  \be
 K_1 (E) = T_{31} + T_{41}; \qquad K_2(E) = T_{23} + T_{24} . \label{4}
  \ee
  \be
 R(E) = T_{33} + T_{34} + T_{43} + T_{44}, \label{5}
  \ee
  The matrix elements  $T_{jk}$ introduced in the Eq. (\ref{2}) are closely related to the elements of  the scattering matrix $S$ describing the whole system $ (T_{jk} = |S_{jk}|^2).$ The latter expresses outgoing from the system wave amplitudes 
$b_L', b_R', a_3', a_4'$ in terms of the incident ones  $b_L, b_R, a_3, a_4.$ The matrix elements $S_{jk}$ are computed below. To reach the middle site in the molecular chain, an injected electron must first tunnel from the left electrode to the chain, and then it tunnels through $ m-1$ barriers separating the sites. The tunneling through this set of barriers may be described by the $2\times 2$ matrix $ W_L$ whose elements are expressed in terms of transmission and reflection amplitudes for the barriers:
    \be  
  W_L = 
\left( \begin{array}{lr}
u_m^L &-v_m^L\\
v_m^L & u_M^L \\
\end{array} \right); \qquad (u_m^L)^2 + (v_m^L)^2 =1. \label{6}
  \ee
  
  Assuming all barriers separating the sites to be identical, we may write the following expression for this matrix:
  \be
 W_L = 
\left( \begin{array}{cc}
\ds\frac{1}{t^L} &\ds-\frac{r^L}{t^L} \\
\ds\frac{r^L}{t^L} & \ds\frac{1}{t^L} \\ 
 \end{array} \right)
\left( \begin{array}{cc}
\ds\frac{1}{t} &\ds-\frac{r}{t} \\
\ds\frac{r}{t} & \ds\frac{1}{t} \\
 \end{array} \right)^{m-1}   \label{7}
 \ee
   where $t^L,r^L, t$ and $r$ are the transmission and reflection coefficients for the barrier between the left electrode and the chain, and for a single barrier separating the sites on the left half of the chain, respectively. The electron tunneling through the right half of the chain could be described in a similar way by using the matrix $ W_R:$
  \be
 W_R = 
\left( \begin{array}{lr}
u_m^R & -v_M^R \\ v_m^R & u_M^R \\
 \end{array} \right) \equiv
\left( \begin{array}{cc}
\ds\frac{1}{t} &\ds-\frac{r}{t} \\
\ds\frac{r}{t} & \ds\frac{1}{t} \\
 \end{array} \right)^{m-1}
\left( \begin{array}{cc}
\ds\frac{1}{t^R} &\ds-\frac{r^R}{t^R} \\
\ds\frac{r^R}{t^R} & \ds\frac{1}{t^R} \\
 \end{array} \right)                \label{8}
 \ee
  where the transmission $(t^R)$ and reflection $(r^R)$ coefficients characterize the barrier between the chain and the right electrode.

 Combining Eqs. (\ref{1}), (\ref{6}) and (\ref{8}) we obtain the following expression for the scattering matrix  $S:$ 
   \begin{widetext} \be
 S = \frac{1}{Z}
\left( \begin{array}{cccc}
u_m^L v_m^R +\alpha^2 u_m^R v_m^L & \alpha & u_m^R \beta & \alpha\beta v_m^R \\
\alpha & u_m^R v_m^L  + \alpha^2 u_m^L v_m^R & \alpha\beta v_m^L & u_m^L\beta \\
\beta u_m^R  & \alpha\beta v_m^R & \beta^2 u_m^R v_m^L & \alpha(v_m^L v_m^R - u_m^L u_m^R) \\
\alpha\beta v_m^R & \beta u_m^L  & \alpha(v_m^L v_m^R - u_m^L u_m^R) & \beta^2 u_m^L 
v_m^R\\
\end{array} \right). \label{9}
    \ee \end{widetext}
  Here, $Z = u_m^L u_m^R - \alpha^2 v_m^L v_m^R;\ \alpha = \sqrt{1 -\epsilon};\ \beta = \sqrt\epsilon. $ For a single site bridge symmetrically coupled to the electrodes $ (m=1,\ t^L = t^R \equiv t';\ r^L = r^R \equiv r')\ \ u_m = 1/t', \ v_m = -r'/t',$ and the Eq. (\ref{9}) agrees with the corresponding result of Ref. \cite{28}. Inserting the expressions for matrix elements $ T_{jk} $ determined by Eq. (\ref{9}) into the Eq. (\ref{3}), we may present the electron transmission in the form:

   \be
 T(E) = \frac{(1+\alpha^2) \big[w_m^2 - \alpha^2(w_m^2 - 1)\big]}{2\big[w_m^2 
+ \alpha^2(w_m^2 - 1)\big]^2} \label{10}
  \ee
  where $w_m$ could be expressed in terms of matrix elements of the matrices $W_{L,R}.$ 

Our next step is to find the relation of the quantities $w_m$ to the relevant Green's functions. Disregarding for a while the effects of nuclear motions, we may write the following effective Hamiltonian for the molecular bridge:
  \be
 H_{eff} = H_0 + H_1 + H_L + H_R.  \label{11}
 \ee
 Here, the first two terms describe the bridge itself. Within the adopted approximation their nonzero matrix elements between states $|k\big>$ and $|l\big>$ corresponding to the $kth $ and $lth$ sites on the molecular chain have the form $(H_0)_{kl} = E_0 \delta_{kl} $ and $(H_1)_{kl} = \beta(\delta_{k+1,l} + \delta_{k-1,l}).$ The remaining terms represent self-energy corrections arising due to the coupling of the leads to the chain. Supposing that only the first and the last sites of the molecular bridge are directly coupled to the corresponding leads, we can write
  \be
 (H_L)_{kl} = \delta_{kl} \delta_{k1} \Sigma_L; \qquad
(H_R)_{kl} = \delta_{kl} \delta_{k,2m-1} \Sigma_R.  \label{12}
 \ee
 The self-energy terms could be written in the form:
  \be
  (\Sigma_{L,R})_{kk} = \sum_r \frac{|\tau_{kr}^{L,R}|^2}{E - \epsilon_r + i \sigma}. \label{13}
    \ee
  In the expression (\ref{13}) summation is carried out over the states on the left/right lead, $\epsilon_r $ is the single electron energy for electron state $r $ in the lead, $\tau $ characterizes the coupling strength between the lead and the molecular chain, and $ \sigma $ is a positive infinitesimal parameter.

The electron transmission for the elastic coherent transport through the considered junction is presented as:
   \be
 T= \Gamma_L \Gamma_R|G_{1,2m-1}|^2   \label{14}
  \ee
 where $\Gamma_{L,R} = - 2 \mbox{Im} (\Sigma_{L,R})$, and the relevant Green's function matrix element is given by:
  \be
 G_{1,2m-1} = \big< 1|E - H_{eff}| 2m -1 \big>.  \label{15}
  \ee

 Further,  we proceed within the wide band approximation  for the leads. Within this approximation self-energy terms  (\ref{13}) do not depend on the electron tunnel energy $E.$ Also, we disregard the real parts of $ \Sigma_{L,R} .$
 Then the expression for the Green's function may be obtained in the form  described in the Ref. \cite{27}, namely:
    \begin{align}
& \!\!\!\! G_{1,2m-1} (E)
\nn\\
   = & \frac{4^m \beta^{2(m-1)} \zeta}{(\lambda + \zeta)^{2(m-1)} (\lambda + \zeta +i\Gamma)^2 - (\lambda - \zeta)^{2(m-1)} (\lambda - \zeta +i\Gamma)^2}.
  \label{16} 
 \end{align}
   Here, we introduced denotations:
  \be
 \lambda = E_0 - E; \qquad \zeta = \sqrt{\lambda^2 - 4\beta^2}; \qquad \Gamma = \Gamma_L + \Gamma_R.  \label{17}
  \ee
In the absence of phonons $(\alpha =1),$ the expression (\ref{10}) for the electron transmission takes on the form:
  \be 
 T(E) = \frac{1}{(2 w_m^2 - 1)^2} \equiv \frac{\theta_m^4}{(2 - \theta_m^2)^2}
            \label{18}  \ee
  where $\theta_m^2 = 1/w_m^2. $ Comparing this result with Eq. (\ref{14}) we obtain:
 \be 
 \theta_m^2 = \frac{2g_m}{1 + g_m}  \label{19}
 \ee
  where the function $ g_m(E) $ is closely related to the Green's function of the chain, namely:
  \be
 g_m(E) = \sqrt{\Gamma_L \Gamma_R} \big  |G_{1,2m-1} (E)\big|.  \label{20}
 \ee
  Using the result of (\ref{19}) we may rewrite the general expression for the electron transmission (\ref{10}) as follows:
  \be
 T(E) = \frac{g_m(E) (1 +\alpha^2) \big[1+ g_m(E) - \alpha^2 (1 - g_m(E))\big]}{\big[1 + g_m(E) + \alpha^2(1 - g_m(E)) \big]^2}.   \label{21}
  \ee
 This expression is used as a starting point in the further analysis. 

\section{iii. results and discussion}

In Fig. 2, we show the energy dependencies of $T(E)$ calculated assuming the parameter $\epsilon $ to be  independent of the energy $E.$ As expected, instead of a single maximum at $E = E_0$ we get a set of peaks located within the interval $ E_0 - 2\beta < E <E_0 + 2\beta. $ The number of peaks equals to the number of sites in the chain. These peaks correspond to the energy levels of the molecular bridge within the adopted model. The coupling of the chain ends to the leads affects the minimum values of transmission, especially near $ E = E_0. $ As the coupling strengthens, the transmission minimum values increase. When the reservoir is attached to the chain $(\epsilon >0),$ it causes a decrease in the range of the transmission variations. However even within the strong dephasing/inelastic limit $(\epsilon =1)$ the transmission peaks remain well distinguishable for multisite molecular bridges. On the contrary, the peak in the transmission through the single site bridge appears completely washed out under strong dephasing \cite{29}.

\begin{figure}[t] 
\begin{center}
\includegraphics[width=7.5cm,height=5.cm]{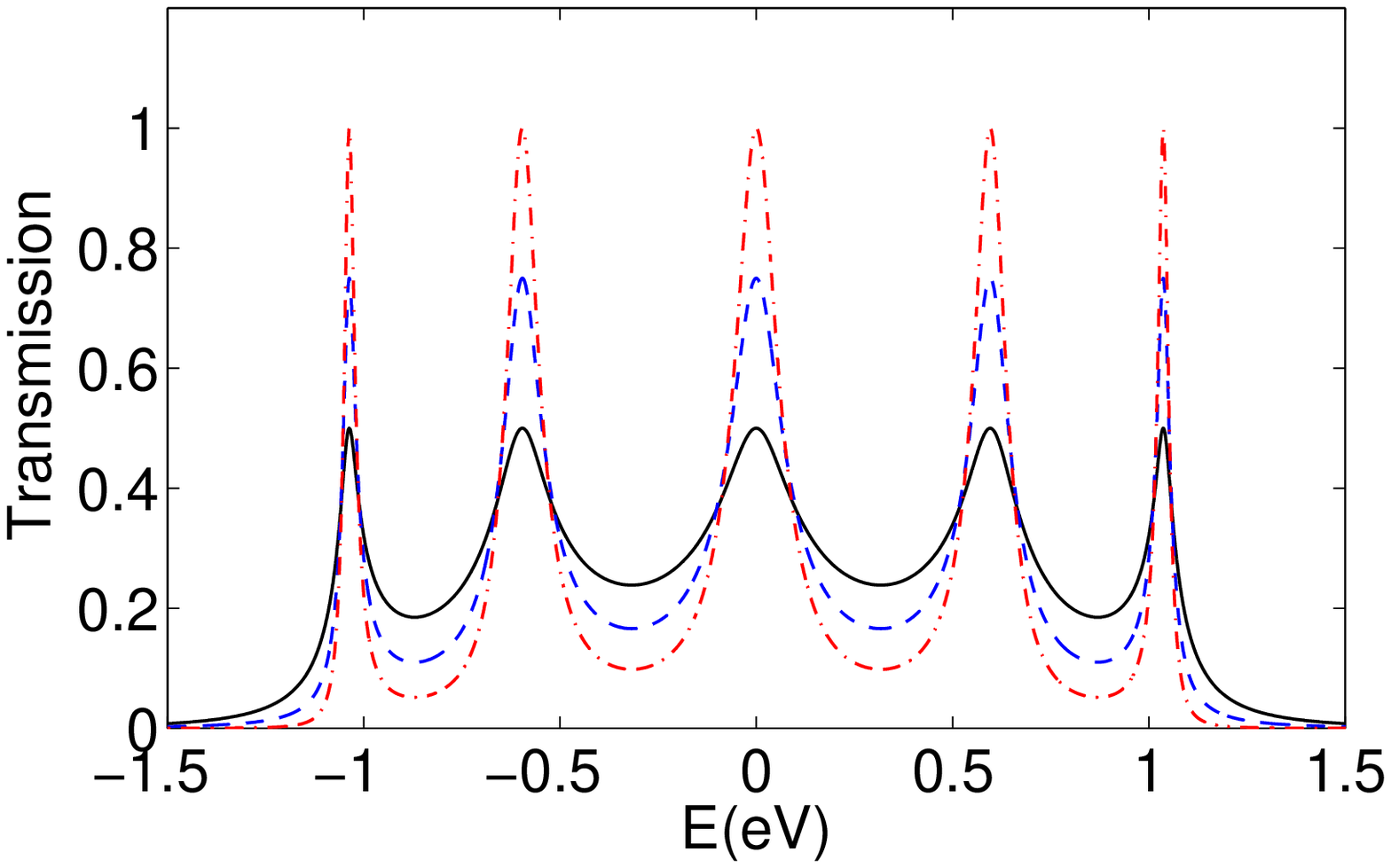}
\includegraphics[width=7.5cm,height=5.cm]{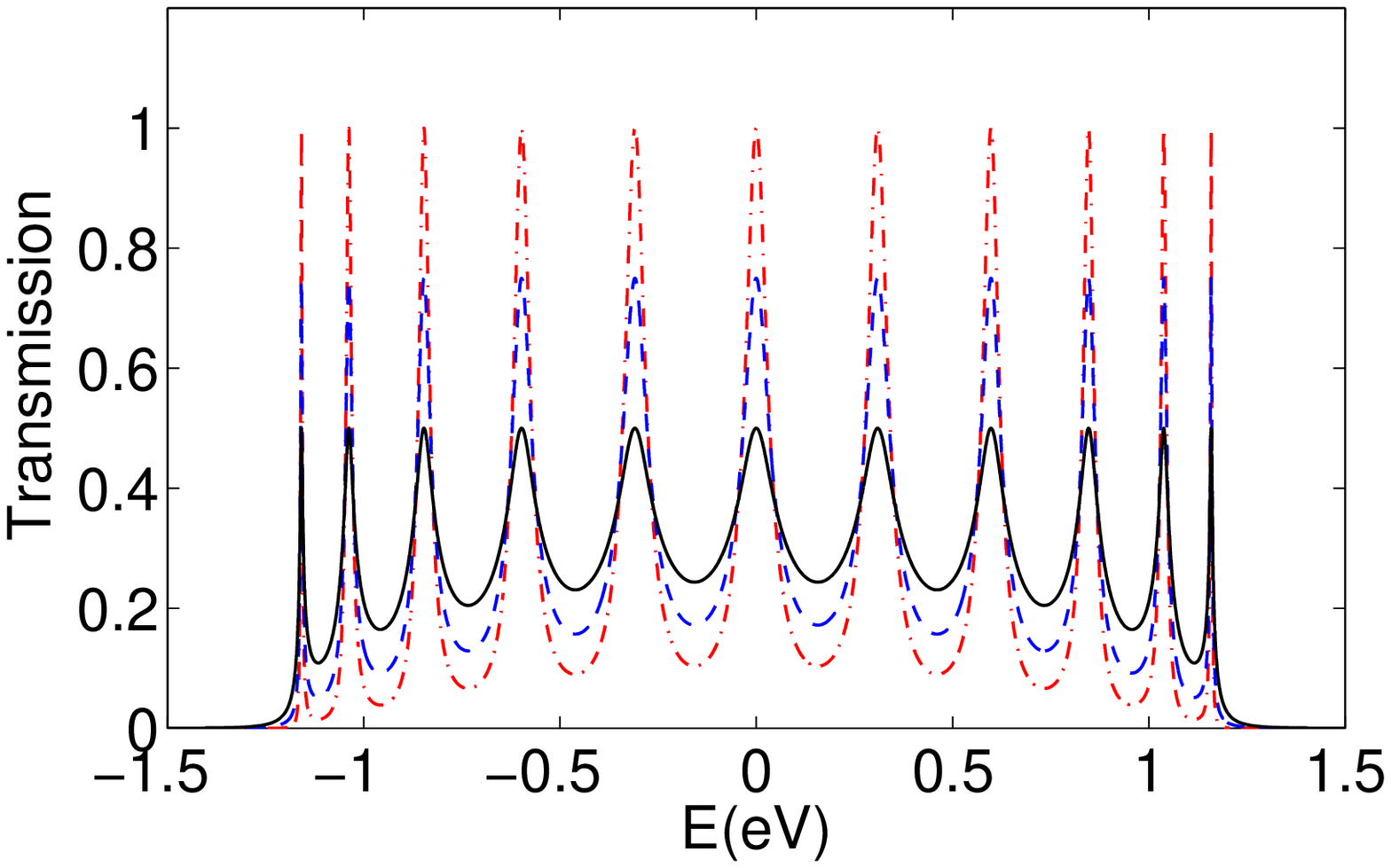}
\includegraphics[width=7.5cm,height=5.cm]{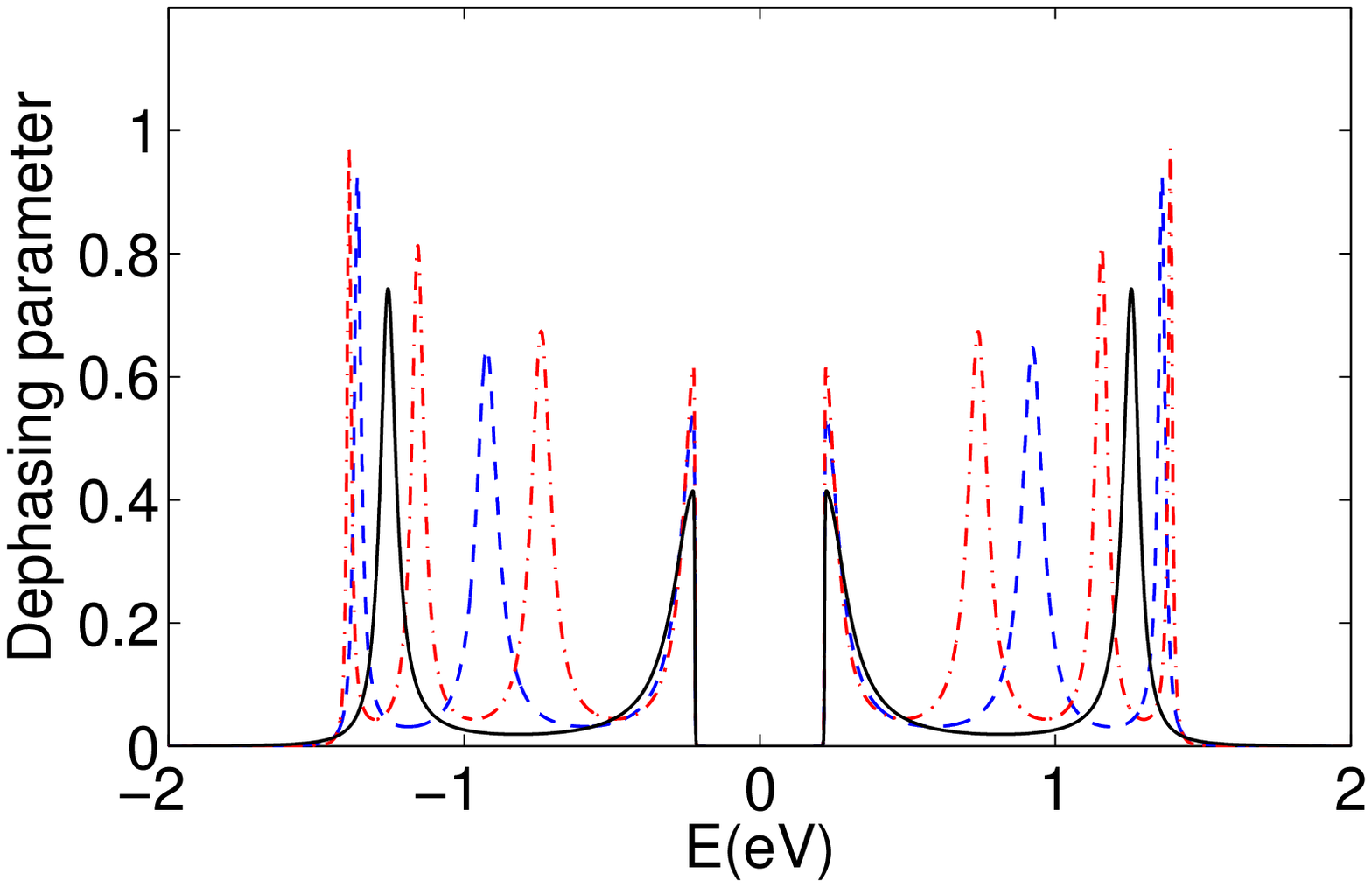}
\end{center}
\caption{(Color online) Energy dependencies of the electron transmission through the junction (top and middle panel) and the parameter $\epsilon $ (bottom panel). The curves are plotted assuming $\epsilon = 0$ (dash-dotted lines), $ \epsilon = 0.5 $ (dashed lines), and $ \epsilon = 1 $ (solid lines) for $ N = 5 $ (top panel) and $N=11$ (middle panel). The curves in the bottom panel are plotted at $N = 11$ (dash-dotted line), $N=9$ (dashed line), $N = 7 $ (solid line) for $M = 0.3 eV,\ \hbar\Omega = 0.22 eV.$  In plotting all curves included in the figure it is assumed that $ \Gamma_L = \Gamma_R = 0.2eV,\ \beta = 0.6eV,\ E_0 =0. $}
\label{rateI}
\end{figure}

  In the considered case of the vibration-induced inelastic contribution to the molecular conductance, the parameter $ \epsilon $ must depend on the quantity  $\Sigma_{ph}(E),$ which is the self-energy correction arising due to the electron-phonon interaction. Comparing the results for the electron transmission obtained basing on the phenomenological Buttiker's model with those derived within NEGF formalism, it was shown that for a single state bridge $ \epsilon $ is rather simply related to $\Sigma_{ph},$namely \cite{29}:
  \be 
 \epsilon = \frac{\Gamma_{ph}}{\Gamma + \Gamma_{ph}}  \label{22}
 \ee
 where $\Gamma_{ph}= - 2\mbox{Im}(\Sigma_{ph}),$ and $\Gamma $ describes the coupling of the bridge to the leads. Presently, we are considering a multisite bridge but we assume that only one site in the middle of the latter is coupled to the vibrational mode. Therefore we approximate $\epsilon $ by the expression similar to Eq. (\ref{22}) where the term $\Gamma $in the denominator is replaced by the energy $ \Delta $, which depends on both  the bridge ends coupling to the leads $(\Gamma)$ and site to site couplings on the bridge itself $(\beta)$. For sufficiently long molecular bridges $(N>10)$ one may expect the value of $ \Delta $ to be mostly determined by the coupling between the sites.

  The form of $\Gamma_{ph}$ varies depending on the characteristics of the electron-phonon interaction in the system, and on the kind of phonons contributing to these interactions. For instance, the electron coupling to the thermal phonon bath results in the expression for the $\Gamma_{ph}$, which strongly differs from that obtained while considering the electron coupling to a few vibrational phonon modes.
   Here, we assume that a single vibrational mode with the frequency $\Omega$ exists in the system and that the temperature is low $(kT \ll \hbar\Omega,\Gamma,\beta).$ 
   Then $\Gamma_{ph} $ could be presented in the form \cite{7}:
  \begin{align}
 \Gamma_{ph}(E) = & \pi M^2 \left\{\int_0^{(\mu_L -E)/\hbar} d\omega \rho_{ph}(\omega) \rho_{el}(E + \hbar\omega) \right.
  \nn\\  & + \left.
\int_0^{(E- \mu_R)/\hbar} d\omega \rho_{ph}(\omega) \rho_{el}(E -\hbar\omega) \right \}
      \label{23}       \end{align}
  where $M$ is the electron-phonon coupling strength, $\mu_{L,R} $ are the chemical potentials for the leads, and $ \mu_R <E<\mu_L.$ 
 The phonon density of states $\rho_{ph} (\omega)$ included in this expression reveals a sharp maximum at $\omega = \Omega:$
  \be
\rho_{ph} (\omega) = \frac{1}{\pi\hbar} \frac{\gamma}{(\omega- \Omega)^2 + \gamma^2}.                 \label{24}   \ee
  Here, the linewidth $ \hbar\gamma $ is of the order of the thermal energy $ kT. $
    The electron density of states (DOS) of the middle site of the chain is described by the expression $\rho_{el}(E) = \ds -\frac{1}{\pi}\mbox{Im} [G_{mm} (E)] .$ Within the adopted model we obtain:
  \begin{align}
 & G_{mm}(E) = \frac{1}{\zeta}
  \nn\\  & \ \times 
 \frac{(\lambda + \zeta)^{m-1} (\lambda+\zeta+i\Gamma) - (\lambda - \zeta)^{m-1} (\lambda - \zeta +i\Gamma)}{(\lambda + \zeta)^{m-1} (\lambda+\zeta+i\Gamma) + (\lambda - \zeta)^{m-1} (\lambda - \zeta +i\Gamma)}.   \label{25}
  \end{align}

 Now, we may apply the obtained approximation for the parameter $ \epsilon $ to compute the vibration-induced electron transmission, and then we employ the latter to calculate the  electron current through the junction. We accept the usual Landauer expression for the current:
 \be 
 I= \frac{e}{\pi\hbar} \int T(E)[f_L(E) - f_R(E)] dE  \label{26}
 \ee
 where $f_{L,R}(E)$ are Fermi distribution functions for the leads with chemical potentials $\mu_{L,R}.$ For a symmetrically coupled junction $\mu_{L,R} = E_F \pm \frac{1}{2}V,$ where $ E_F$ is an equilibrium Fermi energy of the junction. The current given by Eq. (\ref{26}) includes the vibration-induced inelastic contribution. The latter is brought in there by means of the expression (\ref{21}) for the electron transmission $T(E).$ As shown above, the parameter $\alpha^2 = 1 - \epsilon $ in the Eq. (\ref{21}) depends on the electron-vibron interaction, and this leads to the occurrence of the inelastic term in the current along with the elastic one. The proposed formalism enables us to qualitatively  analyze the effects of the electron-vibron interactions in the IETS for multisite molecular bridges.

 The electron-vibron interaction caused self-energy $ \Gamma_{ph} $ strongly depends on the tunnel energy $E,$ and this results in a well pronounced energy dependence of the  parameter $ \epsilon $  shown in the bottom panel i Fig. 2. We remark that $\epsilon$ remains  zero for energies within the range $ E_0 - \hbar\Omega  < E < E_0 + \hbar\Omega. $ This fits into the accepted model where electron-vibron interactions with a single phonon mode are assumed to mainly cause inelastic effects in the electron transport. An electron on the bridge may virtually absorb and emit phonons of the energy $ \hbar\Omega $ thus creating metastable states around the energy states of the molecular chain. In general, the bridge energy states are shifted from their original positions due to the electron-phonon interactions but in the present consideration we disregard these shifts for simplicity. Within our model for the bridge one of the bridge states has the energy $E_0,$ the same as the ionization energy of noninteracting sites. Being initially at this energy level, an electron could make transitions to the states $E =E_0 \pm n\hbar\Omega $as a result of its interaction with the vibrational mode. So, change in the energy  of such electron originating from the interactions with vibrational phonons cannot be less than $\hbar\Omega.$ Otherwise, the electron transport must be coherent and elastic $(\epsilon =0).$ At $E = E_0 \pm \hbar\Omega ,$ the parameter $\epsilon $ sharply increases revealing peaks, which correspond to the appearance of the metastable vibration-induced states around $ E = E_0. $ The same consideration could be applied to another electronic states on the bridge. As expected, $ \epsilon $ versus $ E $ plot displays  a sequence of peaks whose number is related to the number of elements included in the chain. All these features are located in the energy range $E_0 - 2\beta < E < E_0 + 2\beta, $ and $\epsilon $ again becomes zero outside of this range, as shown in the Fig. 2.

\begin{figure}[t] 
\begin{center}
\includegraphics[width=6.cm,height=7.5cm,angle=-90]{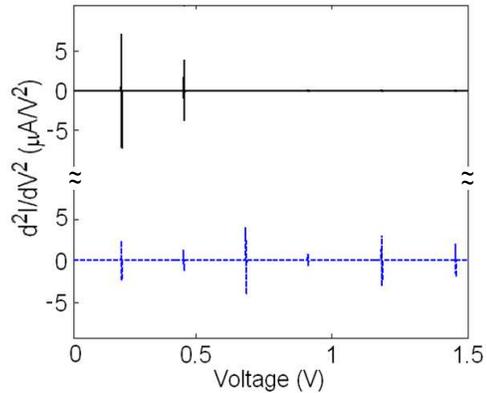}
\end{center}
\caption{(Color online) Inelastic electron spectrum $d^2I/dV$ against $V$ for a molecular junction with a single site (top) and multisite (bottom) bridge. The curves are plotted assuming $ \Gamma_L = \Gamma_R = 0.2 eV,\ \beta = 0.4 eV,\ E_o = 0,\ \hbar\Omega = 0.22 eV,\ M = 0.3 eV, \ N = 7$ (bottom line).  }%
\label{rateI}%
\end{figure}

\begin{figure}[t] 
\begin{center}
\includegraphics[width=4.2cm,height=5cm]{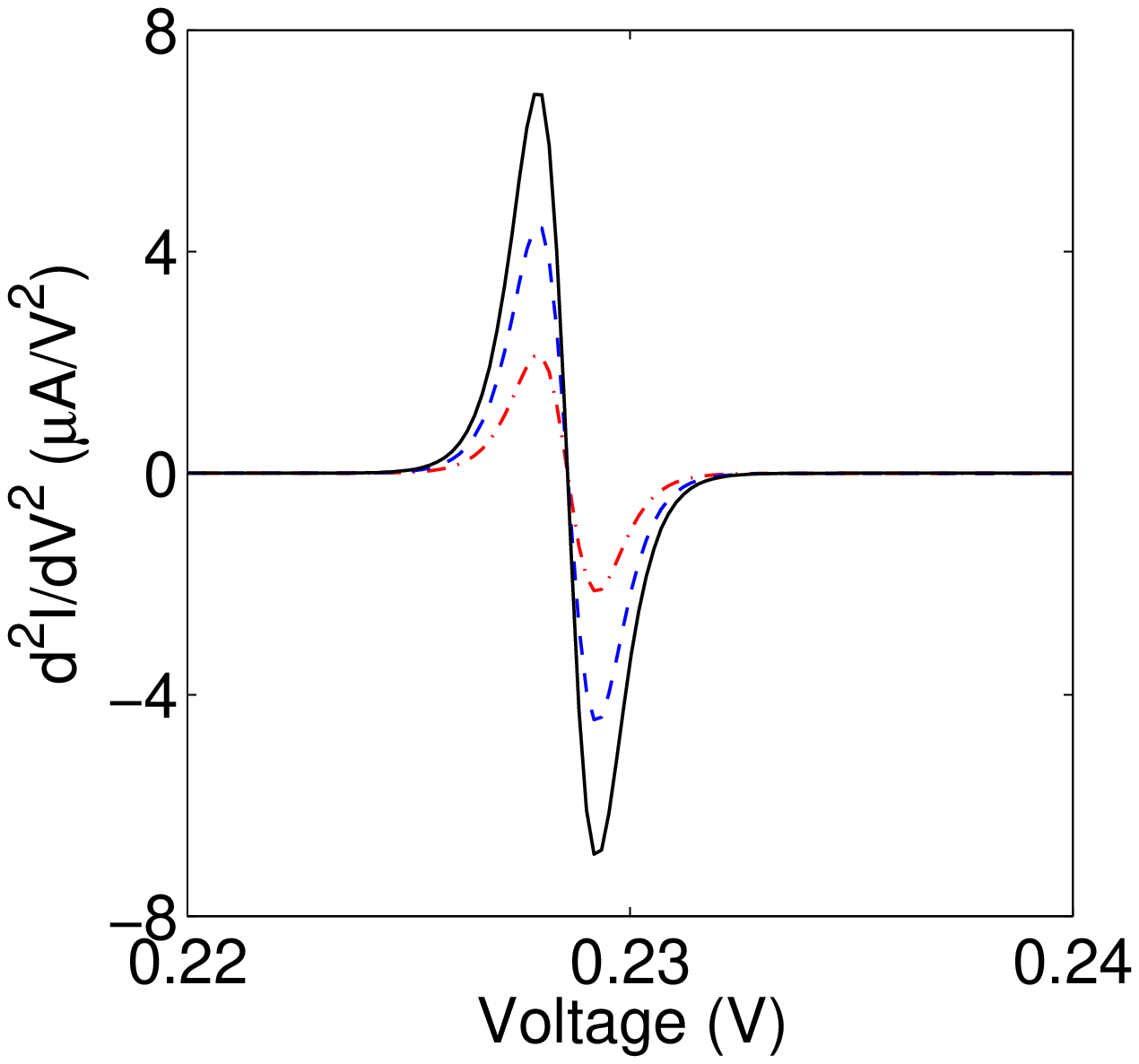}
\includegraphics[width=4.2cm,height=5cm]{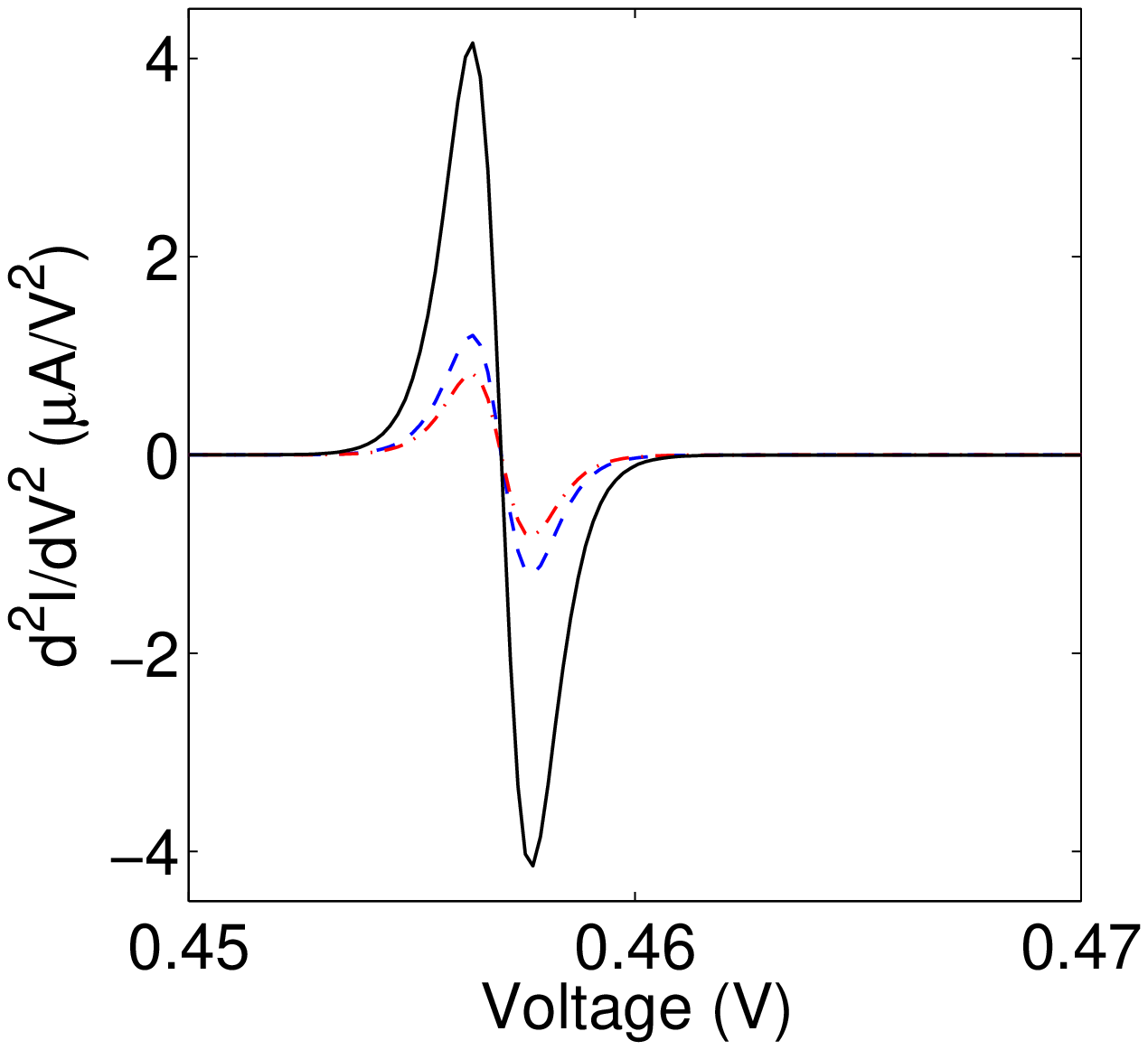}
\includegraphics[width=4.2cm,height=5cm]{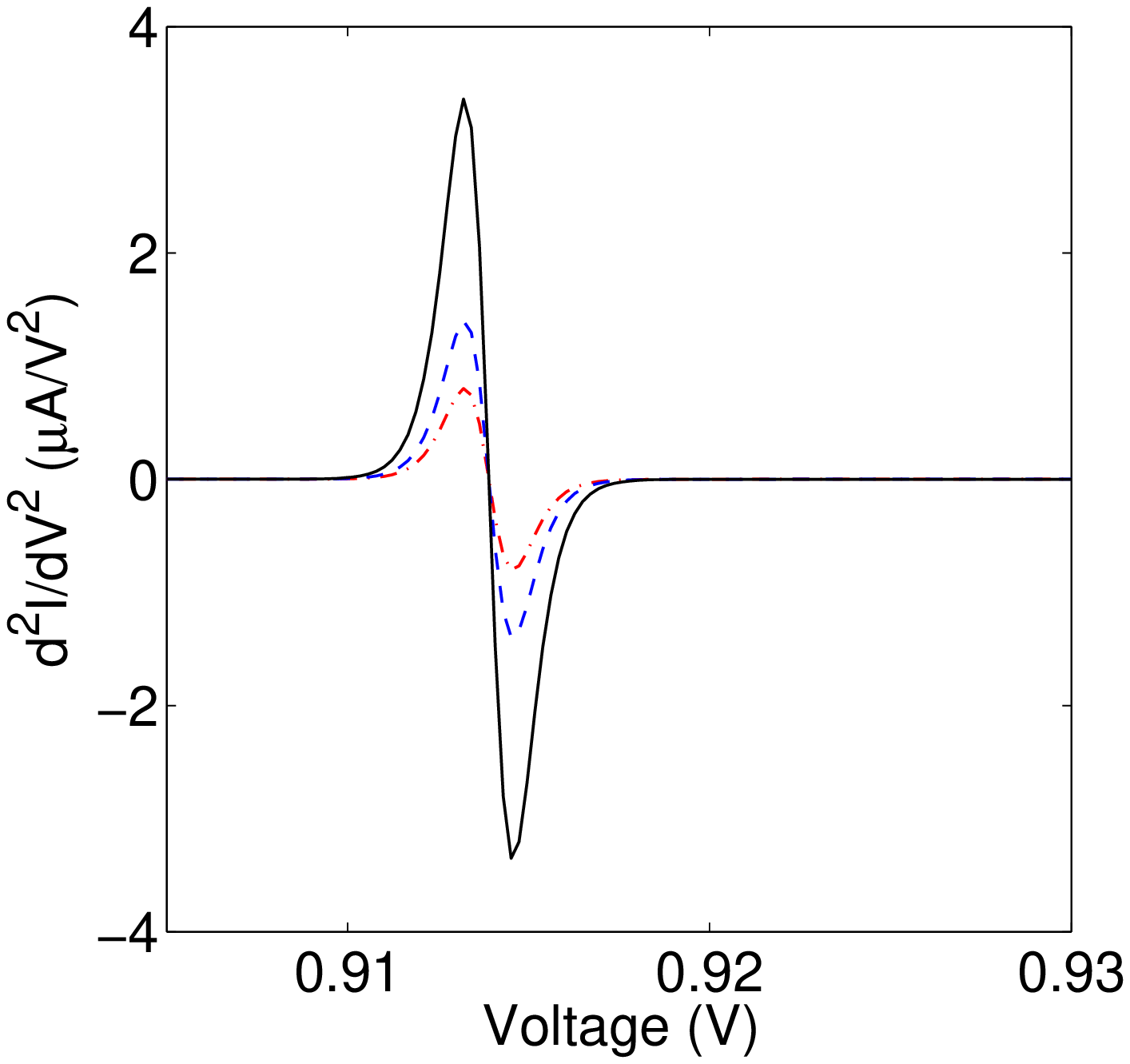}
\includegraphics[width=4.2cm,height=5cm]{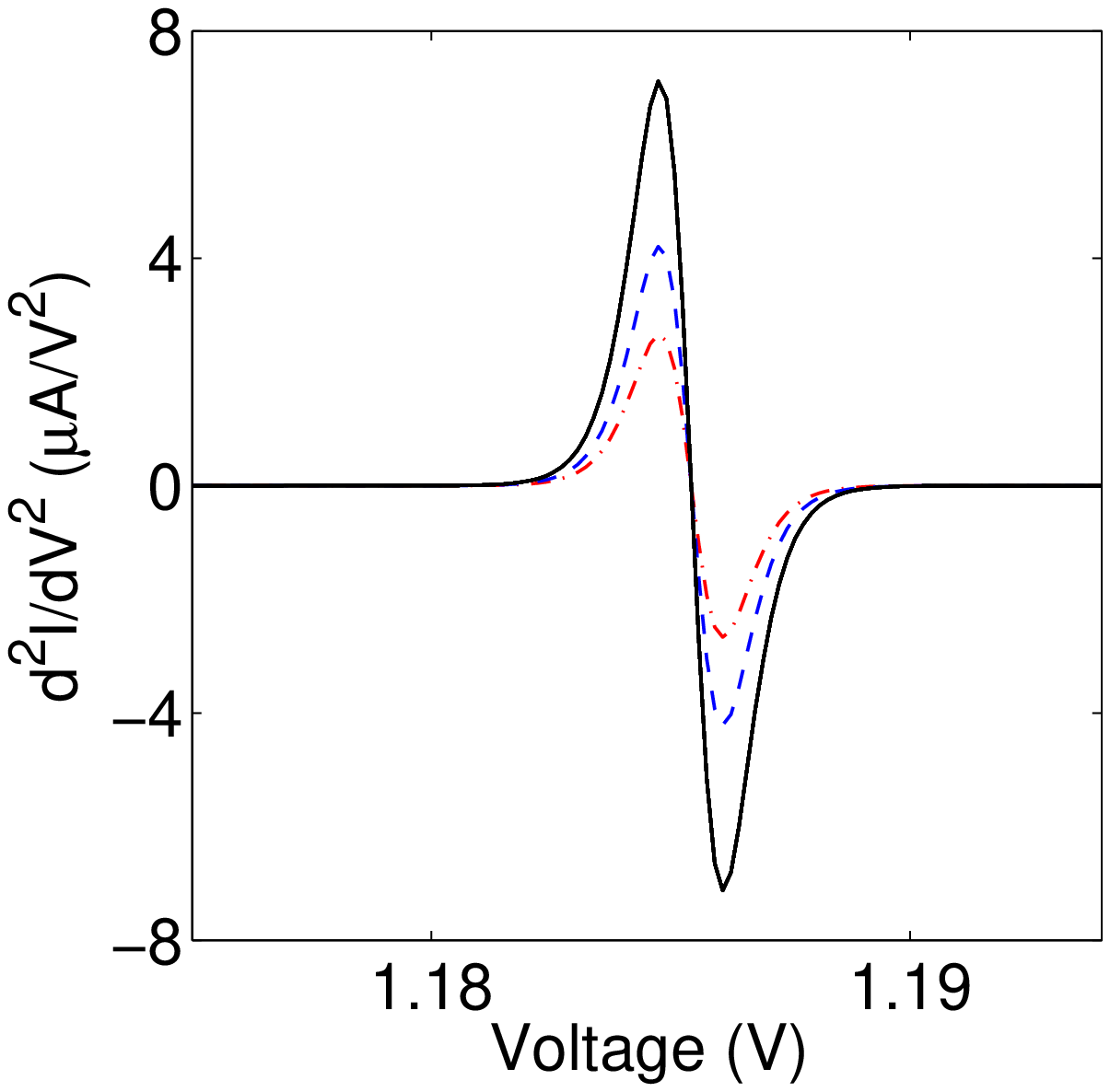}
\end{center}
\caption{(Color online) The IETS features in $ d^2I/dV^2$ for the several resonant levels model (multisite bridge) plotted assuming $\Gamma_L = \Gamma_R = 0.2 eV,\ \beta = 0.4 eV,\ \hbar\Omega = 0.22 eV, M = 0.3 eV, E_0 = 0.$ Top panels: $ N=11$ (dash-dotted lines), $N=3$ (dashed lines), $N=1$ (solid lines). Bottom panels: $ N=11 $ (dash-dotted lines), $N=7$ (dashed lines), $N=3$ (solid lines). }%
\label{rateI}%
\end{figure}

Due to the interaction with the phonon mode, the electron transmission gets extra features which are revealed in the observable characteristics of the electron current through molecular junctions. The specifics of these manifestations is determined by the relative values of the relevant energies. At low temperatures $(kT \ll \Gamma_{L,R}, \beta, M)$, the relation of the bridge-to leads coupling strengths $ \Gamma_{L,R} $ and electron-vibron coupling constant $M $ is especially important. At weak coupling of the bridge to the leads, the electron-vibron interaction splits every single peak in the coherent electron transmission (located, for instance, at $E = \tilde E)$ into a set of smaller peaks associated with vibrational levels at $ E = \tilde E + n\hbar \Omega $ which could be resolved when $\hbar\Omega < \Gamma_{L,R},\beta. $ Phonon-induced peaks in the transmission could give rise to the steps in the $ I-V$ curves and rather sharp features in the differential conductance and IETS plotted against the bias voltage applied to the junction. This was first theoretically shown by Wingreen et al \cite{30} and then confirmed in some other works (see Refs. \cite{7,31}) for the case of a single state bridge. Here, we concentrate on another situation, assuming that the bridge coupling to the leads is rather strong: $\Gamma_{L,R} \sim M.$ In this case inelastic effects       do not cause significant changes in the $I-V $ characteristics.  However they cause the appearance of a distinct signal in the inelastic tunneling spectrum at $ eV \approx \hbar\Omega_0 $ similar to that occurring on a single-site bridge interacting with the vibrational mode. The shape and magnitude of the signal are sensitive to the characteristics of the considered system.

\begin{figure}[t]  
\begin{center}
\includegraphics[width=5.6cm,height=5.2cm]{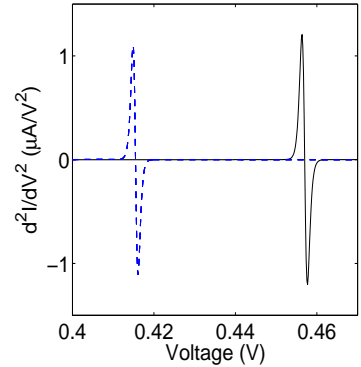}
\end{center}
\caption{(Color online) The effect of asymmetric coupling of the bridge to the leads on the IETS feature in $ d^2I/dV^2.$ The curves are plotted for $N=7,\ \Gamma_L=\Gamma_R=0.2eV$ (solid line) and $\Gamma_L = 0.18 eV,\ \Gamma_R = 0.22 eV $ (dashed line).  The remaining parameters take on the same values as in the Fig. 4.}%
\label{rateI}%
\end{figure}

  We remark that our semiphenological approach brings the result which agrees with those earlier obtained in several theoretical works basing on the consistent NEGF for single state molecular bridges. The size of the derivative -like IETS signal at $eV \approx \hbar\Omega $ depends on the number of sites in the bridge chain as shown in the figures $3 $ and $ 4 $ (top left panel). For $ N=1 $ (a single-site bridge), the signal size takes on value of the same order as that reported by Galperin et al \cite{7} for a symmetrically coupled junction with relatively close values of the relevant parameters. For multisite bridges the IETS reveals more signals which indicate inelastic electron transport through the channels associated with the bridge energy levels situated apart from $ E=E_0.$ An example of the vibron-induced IETS  for a multisite bridge $(N=7)$ is presented in the right panel of Fig. 3. Some IETS signals are separately displayed in Fig. 4 to show their shape and size for different values of $N.$

Our analysis shows that the signal indicating the phonon excitation threshold is accompanied by another feature appearing at $eV \approx 2\hbar\Omega, $ and this happens for multisite and  single-site bridges in the same way. The multiplication  of IETS signals is known for molecular junctions with weak coupling of the bridge to the leads, and it shows the contribution of higher phonon harmonics into the electron transport. However, this explanation is hardly justified for the  presently considered junction with rather strong coupling of the bridge to the leads. To clarify the nature of the obtained IETS signal duplication, we computed $d^2 I/dV^2$ for a junction with a slightly asymmetric coupling. It appears that the extra signal position varies depending on the ratio $\Gamma_L/\Gamma_R, $ (see Fig. 5), and the higher is the asymmetry, the closer it moves to the phonon excitation threshold. In general, the extra signal appears at $(1 + \Gamma_L/\Gamma_R) eV \approx \hbar\Omega.$ One may expect that within the scanning tunneling microscopy (STM) junction configuration $(\Gamma_L \ll \Gamma_R)$ this signal would be superimposed on that indicating the excitation threshold for the vibrational phonons. This gives grounds to conjecture that the effect of the electron-vibron interactions on the IETS is twofold. The very opening of the channel for inelastic transport gives rise to the signal in the inelastic tunneling spectrum of electrons. This signal always occurs at the phonon excitation threshold regardless of the characteristics of the bridge coupling to the leads. Another signal indicates that the corresponding metastable electronic state appears in the conduction window between $\mu_L$ and $\mu_R$, and its position is determined by the bias voltage distribution in the junction. The two coincide within the STM configuration, and their maximum separation happens in symmetrically coupled junctions and equals $\hbar\Omega. $ Similar signal duplication arises due to the opening of inelastic transport channels associated with every electronic state of the bridge chain.

Finally, we remark that the adopted computational approach proved itself workable. It allows us to reproduce the principal vibration-induced inelastic effects in the electron transport through molecular junctions and to save computational efforts and time. In the particular case of a single-site bridge linking the leads in the junction, the proposed approach gives results which agree with those obtained using advanced computational formalisms. At the same time, the adopted model enables us to analyze inelastic electron transport through multisite bridges simulating practical molecules. One may generalize the proposed computational method to include into consideration a set of vibrons with different energies. For this purpose one must include into the model extra dephasing reservoirs attached to the bridge sites. Also, electron structure of a practical molecule may be used in calculations of the bridge Green's functions, and this could help to further improve the results, and make them more useful and convenient for comparison with experiments.

\section{acknowledgements}
We  thank M. A. Ratner and V. Mujica for stimulating and helpful discussions and  G.M. Zimbovsky for his help with manuscript.  NZ gladly acknowledges support from NSF-DMR-PREM 0353730 and DoD grant W911NF-06-1-0519.

\end{document}